# The comparison of the healing performance of polyurethane-modified bitumen mixtures

Mohammadjavad Kazemi[1], Ahmad Goli, Abbas Mohammadi, Mohsen Aboutalebi Esfahani

**Abstract**:

Considering much distresses occurring at the surface of the asphalt roads, self-healing phenomenon attracted much attention because of reducing the road surface distresses and its benefits. The use of polymers, which are mainly used to improve the properties of asphalt mixtures, is one of the ways to increase self-healing properties. In this study, four polymer-modified bitumen mixtures were made applying most common polyurethanes (by combining three different types of polymers and a nanographene particle). Then the mixtures were exposed to long-term aging under PAV to be evaluated in situations where less healing occurs. Finally, by performing a time sweep test and measuring the shear complex modulus (G *) before and after the rest period, the healing rate of these mixtures was measured and compared. By examining the results, it was found that the TDI-CO polymer-modified mixture is the best mixture for bitumen healing.

# 1 Introduction

One of the problems and issues in asphalt mixtures is the occurrence of failures and cracking in them under different conditions, which first appear as small damages in the surface and then expand over time and under different loading and weather conditions[1]. Many studies have been conducted to reduce these failures and many solutions have been proposed, including the addition of modifying materials. Today, one of the methods to reduce distresses and reverse the cracking process is self-healing. This phenomenon, which is inherent in many materials, improves and reduces failures under various conditions and factors[2]. This inherent property in bitumen, with many applications in road construction and construction industry, has also attracted researchers to use this property increasingly in flexible pavements.

[1] *Department of Civil and Transportation Engineering, University of Isfahan, Isfahan, Iran*

Many factors affect this property, and many researches have been conducted to investigate these factors. These factors are divided into two types of internal and external factors: modifying by adding different materials and aging are two of the internal factors affecting this property[1]. Regarding the positive effects of some polymers on increasing the fatigue life or improving the bitumen performance grade, the usage of polymers has increased in the construction industry, but there are many doubts about the effect of polymer addition on self-healing properties so that some polymers reduce and some other increase the healing [3-6]. Today polymers and polyurethane compounds are used in many industries and have been noticed a lot due to material characteristics improvement, in addition to the traditional use as an adhesive material and foam at present lots of new features have been added to them. These polymers in combination with bituminous material have also been used for different purposes.

In this research, intending to use polyurethane polymers to improve self-healing properties and conduct a healing test with the DSR device, we evaluated four types of polyurethane polymer-modified bitumen mixtures subjected to long-term aging.

## 2 Research background

In recent decades, when the topic of materials' self- healing has been proposed in materials and the production of intelligent materials to increase the application efficiency, many experiments have been carried out to determine the self-healing properties of bitumen. Experiments on bitumen fatigue are divided into two categories including tests based on fatigue and tests based on failure. Fatigue tests are generally conducted using Dynamic shear rheometer device (DSR) in either controlled stress or controlled strain state and the frequency range of 1.59 to 41 Hertz. To arrange the tests with constant stress and constant strain, stress between 60 to 400 Kilo Pascal and a strain level in the range of 0.3 to 20 percent are applied to the samples. Moreover, to ensure that the damage is caused by fatigue, not due to flow instability, experiments are generally administered in the temperature range of 5 to 25[7, 8].

Aging is one of the factors affecting the properties of bituminous materials[9] that according to the various researchers studies these changes and loss of quality reduces bitumen healing capacity[10, 11]. In this context, Little saw an increase in the healing ability of aged bitumen by adding hydrated lime, which reduces aging in the bitumen[12]. Van de burgh investigated the difference between the healing of samples aged in the laboratory with field aged samples and pointed out that the mixture had a higher healing capacity in laboratory samples but naturally aged samples had a lower healing capacity compared to their unaged mixtures[13].

In 2009 Santagata et al. conducted DSR device healing test and SARA test on 6 bitumen mixtures from various sources to find the relationship between bitumen chemistry and its failure and healing performance. They reported that the microstructures of different types of bitumen affect its functional properties and healing. Meanwhile, all four different parts of bitumen (asphaltene, aromatic, resin and saturate) affects the failure resistance as well, but the oily phase is more effective in the healing performance[14].

Shen and Sutharsan also examined the healing rate of bitumen applying a method called rate of dissipated energy change (RDEC) and using two different types of bitumen (PG64-28, PG70-28) under short-term and long-term aging. In these experiments, they used intermittent loading with short-term rest periods among loadings. Finally, they obtained the healing rate of each type of bitumen in the desired condition using plotting graphs based on the Plateau logarithm value and the rest periods ($\ln PV - RP$) and measuring the slope of these graphs. In their results, they also stated that polymeric bitumen (PG70-28) has had a higher healing rate than that of the base bitumen. Besides, about the effects of temperature and frequency on healing, they acknowledged that temperature affects healing performance approximately linearly so that increasing temperature increases the healing rate, but on the contrary, the frequency increase decreases the healing rate. They also found out that strain level affects healing in an adverse and nonlinear way [15, 16].

Canestrari et al investigated the effects of thixotropy and aging rate as well as the effect of SBS polymer on the healing improvement, stating that the distress of a part is reversible and another part is irreversible. Their experiments showed that thixotropy is a factor affecting fatigue and healing. They also reported that the SBS polymer has increased the amount of healing by reducing the irreversible failures and increasing the thixotropy. Unlike other researches, they stated that aging has no effect on the healing amount and only reduces the loading times to achieve a certain amount of fatigue[17]. Baglieri et al. investigated the healing of polymeric bitumen as well and stated that in SBS modified bitumen, increasing the SBS polymer content to a certain extent, improves healing, and after that, reduces healing [18].

In a study to investigate the healing optimal temperature, Tang et al. investigated the optimum healing temperature with a method based on Newtonian law in viscosity. They administered time sweep and frequency sweep tests on two types of bitumen (aged and unaged), to obtain the bitumen properties and their healing rates. The information obtained from these tests concluded that both aged and unaged bitumen show a behavior similar to Newtonian behavior near their softening temperature that is the optimal healing temperature for the healing of existing fatigue cracks. It was also found that increasing aging increases the optimum healing temperature[19].

Due to the increased attention to polyurethane polymers, new researches have begun on the effect of these polymers on bituminous materials. Singh et al. used this category of polymers to examine the coverage and the improvement of water-resistant property of polyester coating. They observed that the polymer network formed by polyurethane has improved elastic recovery, elasticity, and maintenance of adhesion to lower layer[20].

In 2014, Lee et al researched healing asphalt pavements by synthesizing polyurethane polymer and mixing polyurethane / Silicon carbide. They made polyurethane polymers with different weight percent of liquid diphenylmethane diisocyanate (L-MDI) to Toluene diisocyanate (TDI-80) in their study. They combined it with a mixture of silicon carbide and made a mixture to remove the cracking of asphalt mixtures. They examined the properties of silicon carbide composition unmodified and modified with polyurethane administering different mechanical and chemical experiment. Finally, they measured the time required to reopen traffic on layers paved with polyurethane modified silicon carbide. It was found that the composed pavement could be reopened on passing traffic 90 minutes after implementation. Increasing the weight ratio reduces the time required to watch the mixture and mechanical properties initially decreases and then increases. 6 to 4 weight percent is the best for these polymers [21].

Carrera et al examined the effect of polyurethane polymers on emulsion bitumen properties. They also investigated the rheology, storage and stability capacity of this emulsion with various levels of the prepolymer. In the particles size dispersion test of the emulsion bitumen, 50 weight percent of which was composed of bitumen, it was found that the best content of prepolymer to increase its stability is 1 to 2 percent. However, in the mixtures made with 1 weight percent of prepolymer, there was a high viscosity decrease [22].

## 3 Methodology

### 3.1 Materials and Synthesis of Polymers

In this research, the bitumen produced by Jay Oil Company with a performance grade of PG58-22 has been used to prepare polymeric bitumen mixtures and compare them with the base state (unmodified bitumen). The synthesis process of each of the polyurethane polymers is as follows.

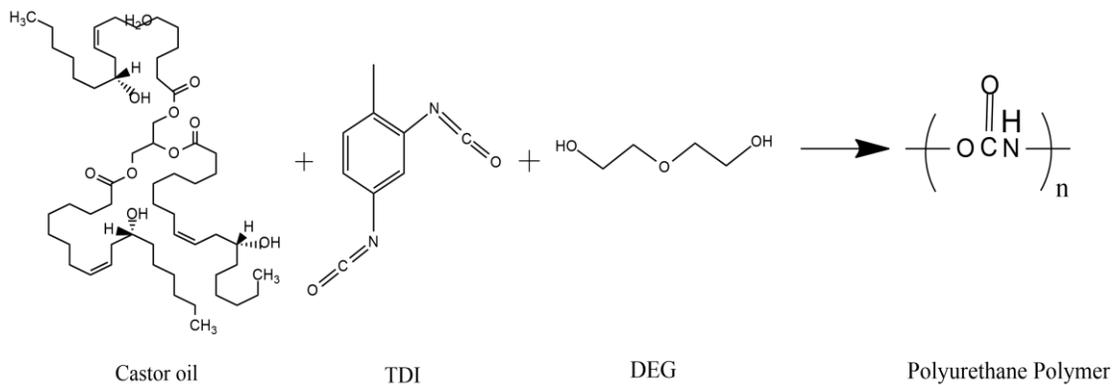

**Figure 1 Chemical structure of materials involve in producing TDI-CO**

To synthesis TDI-CO polymer, the molar ratios of 1: 2: 1 were used for castor oil, toluene diisocyanate (TDI) and diethylene glycol (DEG), respectively (Figure 1). In the prepolymer synthesis process, to remove water from the polyol, it was placed inside the oven with 70 ° C temperature for 24 hours. Then TDI and pre-weighed castor oil (The main prepolymer material) were poured into 3-neck round bottom flask equipped with a nitrogen purge, condenser, and mechanical stirrer. Prepared flasks were placed in an oil bath with 90 ° C temperature and the contents were stirred for 3 hours at 250 rpm. Pre-weighted DEG was added to the mixture at the time of the mixing of prepolymer and bitumen to cure polymer (this produce TDI-CO mixture). In order to produce a mixture of polyurethane and nanographene, the same polymer production process was carried out, and finally, at the time of polymer and bitumen mixing, nanographene was added to the mixture with the 5 wp of the polymer (this produce TDI-CO-GO mixture). The nanographene used was modified by calix[4]arenes to make the nanographene Hydrophobic and this modification was according to Hummers' improved method [23].

To synthesis MDI-CO polymer, the molar ratios of 1: 2: 1 were used for castor oil, methyl diphenyl diisocyanate (MDI) and DEG, respectively (Figure 2). For synthesizing MDI-PPG polymer the molar ratios of 1: 2: 1 were used for PPG, MDI, and DEG, respectively (Figure 3). The synthesis process for these polymers is also like TDI-CO polymer synthesis.

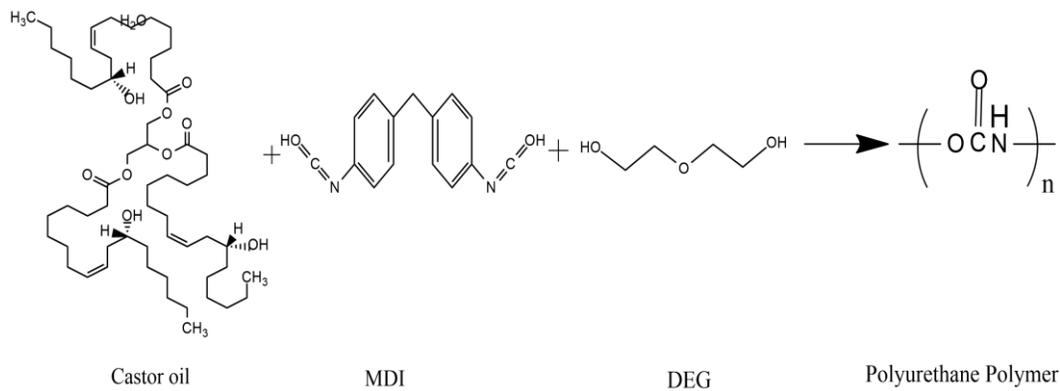

**Figure 2 Chemical structure of materials involve in producing MDI-CO**

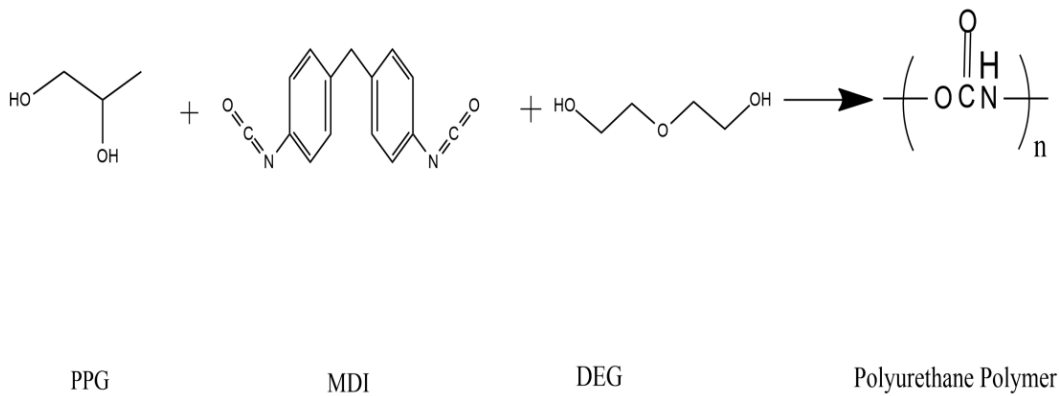

**Figure 3 Chemical structure of materials involve in producing MDI-PPG**

It should be noted that polyurethane polymers that were completely cured did not have proper mixing with bitumen. Therefore, they were prepared as curable polymers (prepolymers) and DEG was added to bitumen at the time of making bituminous mixture. To prepare the polymeric mixture, prepolymer and DEG were added to base bitumen of 140 ° C at the same time to be mixed together using a high-shear mixer for 60 minutes at 2500 rpm. The added polymers was in proportion to 3 weight percent of bitumen.

In line with the previous studies, the amount of healing in aged mixtures was less than unaged mixtures, polymer mixtures prepared according to ASTM D2872 and ASTM D6521 standards were subject to long-term aging [24, 25].

## 3.2 Procedures

After preparation of different mixtures, pure bitumen and bitumen modified with different polymers were poured into silicone molds of 8 Mm diameter and 2 Mm depth and were allowed to cool down. After cooling down, the bitumen poured into the molds were removed out of the mold and placed into the DSR machine. To begin the test, the gap between devise plates was set on 2 mm. During the test, semi sinusoidal loading with controlled strain state and frequency of 10 Hertz and 3% Strain at 20 ° C was applied to samples. During loading samples, shear modulus (G *) was measured and recorded by the device. 100th recorded shear modulus was recorded as the initial shear modulus ($G^*_{initial}$). Loading continued until reaching 70 percent of the sample initial modulus and during this time the loss of modulus was measured. After reaching the target module value, loading has been stopped and samples were put to rest at the same temperature for 1 hour so that in case of healing, shear modulus increases ($G^*_{AR}$). After the rest, the sample was loaded again under the same conditions, so that the shear modulus value reaches the shear modulus value before the rest period ($G^*_{BR}$). To select the desired mixture which is appropriate for healing, each mixture was tested 3 times to ensure repeatability of results. After finishing the tests, the healing index for each of the polymers was calculated by equation 1.

$$HI = \left(\frac{G^*_{AR} - G^*_{BR}}{G^*_{initial} - G^*_{BR}}\right) \times \left(\frac{N_{AR} - N_{BR}}{N_{BR}}\right) \times 100 \tag{1}$$

Where HI is the healing index, $G^*_{initial}$, $G^*_{BR}$ and $G^*_{AR}$ are complex shear modulus at the beginning of the test, the sample modulus before the rest period ( target Module ) and sample modulus after healing period, respectively. $N_{BR}$ and $N_{AR}$ are the number of loading cycles applied on the samples before and after loading to reach a determined amount of loss.

## 4 Results and discussion
### 4.1 Time sweep test

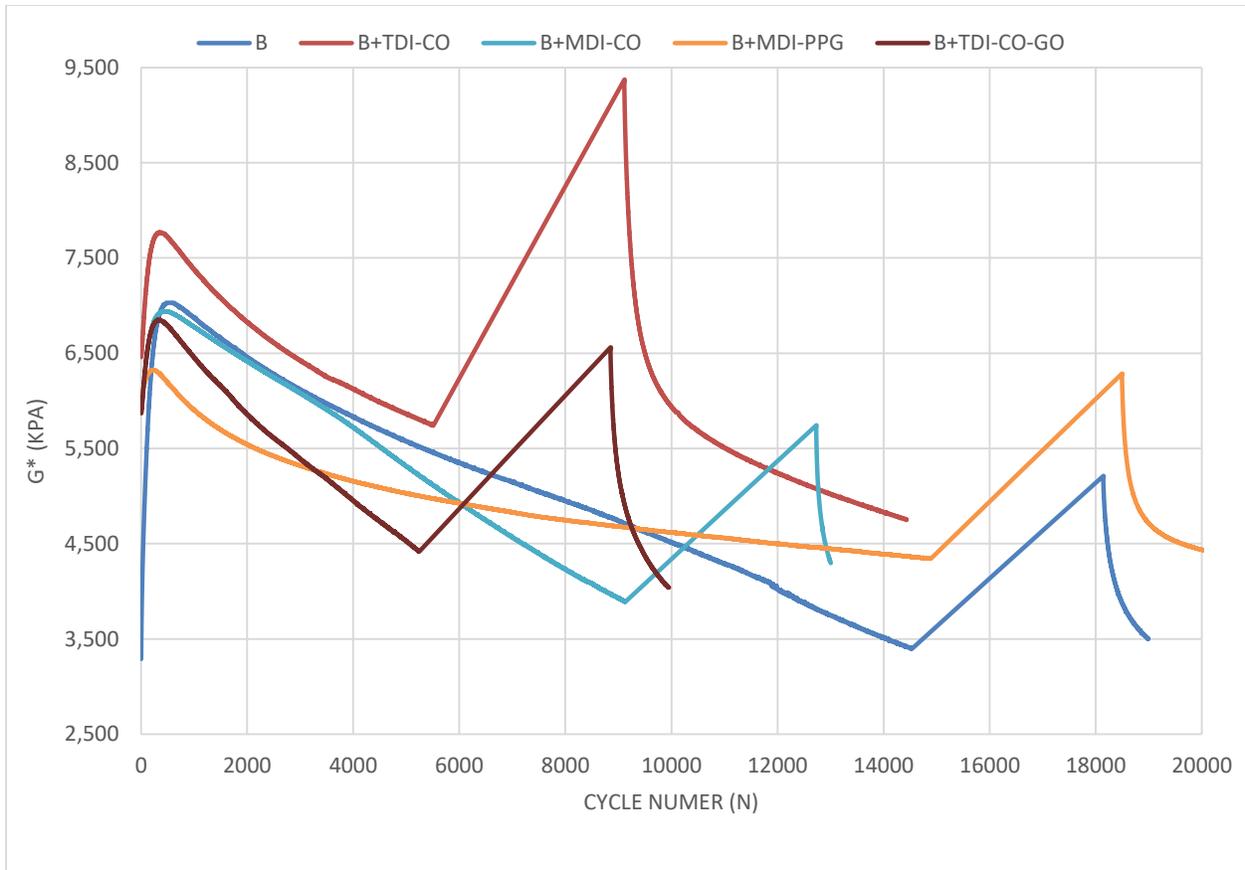

**Figure 4 complex shear modulus chart of bituminous mixtures with different polymers during healing test**

Data obtained from time sweep test on different polymeric bitumen and pure bitumen are presented in figure 4 and table 1, respectively. According to the data, it can be seen that B + MDI-PPG and B have had the highest loading cycle to reach the 30 percent loss in complex shear modulus. B + TDI-CO and B + TDI-CO -GO Mix have had a modulus drop in the lower number of cycles compared to other mixtures, but B + TDI-CO has had the highest rate of modules healing after the one hour healing period. It should also be noted that although the amount of cycle that has borne by B + TDI-CO has been less than other mixtures or pure bitumen, the composition of this polymer has increased the complex modulus of bitumen proportionately. This, in turn, can affect the rheological and functional properties of bitumen and this increase has been much higher than other polymers.

**Table 1 the results of healing tests on the bitumen mixtures with different polymers**

| Mixture type | Mix code | $G^*_{initial}$ (Kpa) | $G^*_{B.R}$ (Kpa) | $G^*_{A.R}$ (Kpa) | $N_{A.R}$ | $N_{B.R}$ | HI | Healing ratio |
|---|---|---|---|---|---|---|---|---|
| Pure bitumen | B | 5370 | 3400 | 5210 | 15514 | 14441 | 6.83% | 1 |
| TDI-CO Polymer modified bitumen | B+ TDI-Co | 7190 | 5740 | 9370 | 6676 | 5410 | 58.58% | 8.6 |
| MDI-CO Polymer modified bitumen | B+ MDI-CO | 6422 | 3889 | 5742 | 9546 | 9029 | 4.19% | 0.6 |
| MDI-PPG Polymer modified bitumen | B+ MDI-PPG | 6239 | 4344 | 6285 | 16892 | 14794 | 14.53% | 2.1 |
| TDI-CO-GO Polymer modified bitumen | B+TDI-CO-GO | 6440 | 4420 | 6560 | 5761 | 5149 | 12.59% | 1.8 |

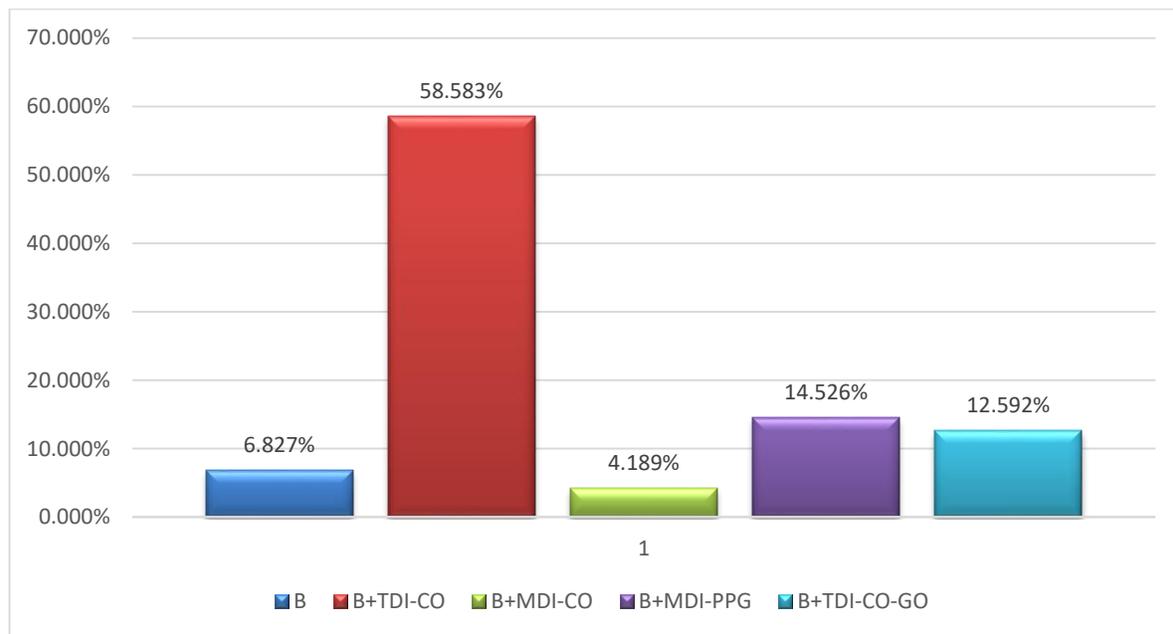

**Figure 5 the Comparison of healing of different mixtures**

According to the healing index of each of the mixtures (figure 5), B + TDI-CO mixture has had the highest healing rate among other mixtures. Healing index of B + MDI-PPG mixture has almost doubled compared to pure bitumen. At the next rank is B + TDI-CO-Go mixture showing the increase of bitumen healing with the composition of TDI polymer made with castor oil, but adding modified nanographene to bitumen has lessened the healing rate. This decrease is probably due to the high surface absorption of this nanoparticle and the adsorption of polymer materials and lumping in the mixture that has reduced the TDI polymer performance. In the last rank is the B+MDI-CO, suggesting that this polymer has disrupted the natural healing process of bitumen and has reduced healing rate. Using castor oil in manufacturing polymers reduces the use of chemicals and replaced them with the use of processed materials from nature, which can be said that these polymers are largely environment-friendly and reduce the use of harmful substances in nature.

Finally, considering the healing rate obtained from B + TDI-CO mixture as well as the environmental impact of castor oil in manufacturing polymer instead of using harmful chemicals, this mixture were chosen as the best polymer among the prepared polymers.

It is noteworthy to say all mixtures had tested three times to ensure the repeatability of test and all same mixtures showed nearly the same result.

## 5   Conclusion

In this study, conducted as a closer examination of the effects of different types of polymers existing in the polyurethane polymers family on self-healing phenomenon, the healing rate was measured by performing a time sweep test using DSR device and three types of polymers and a nanographene particle to improve bitumen. The results are as follows:

- According to the tests conducted to determine the appropriate polymer to increase the healing phenomenon, B + TDI-Co and B + MDI-PPG with the healing of 8.6 and 2.1 times compared to pure bitumen, respectively, have had the highest degree of healing among polymeric bitumen, which means that TDI-CO polymer has effectively increased the self-healing phenomenon.

- The TDI-CO polymer combining with bitumen has increased the complex shear modulus so that after 30 percent loss, it has not yet reached the pure bitumen modulus value. Moreover, in B + MDI-PPG in addition to increasing in healing, there is a relative decrease in the slope of the modulus drop which indicates a decrease in the fatigue rate

- Finally, due to the higher levels of healing and the use of environment friendly materials in the TDI-CO polymer, the bitumen modified with this polymer is the best combination for increasing self-healing and being used in the road construction industry and can be used as a polymer to increase self-healing properties and reduce road failures in field conditions.